\documentclass[12pt]{iopart}

\usepackage{graphicx}
\usepackage{iopams}
\usepackage{color}

\begin{document}

\title[First passage in heterogeneity controlled kinetics: beyond the mean]{First
passage time distribution in heterogeneity controlled kinetics: going beyond the
mean first passage time}

\author{Alja\v{z} Godec$^{\dagger,\ddagger}$ and Ralf Metzler$^{\dagger,\sharp}$}
\address{$\dagger$ Institute of Physics \& Astronomy, University of Potsdam, 14776
Potsdam-Golm, Germany\\
$\ddagger$ National Institute of Chemistry, 1000 Ljubljana, Slovenia\\
$\sharp$ Department of Physics, Tampere University of Technology, FI-33101
Tampere, Finland}

\begin{abstract}
The first passage is a generic concept for quantifying when a random quantity
such as the position of a diffusing molecule or the value of a stock crosses a
preset threshold (target) for the first time. The last decade saw an
enlightening series of new results focusing mostly on the so-called mean and
global first passage time (MFPT and GFPT, respectively)
of such processes. Here we push the understanding of first passage processes a step
further. For a simple heterogeneous system we derive rigorously the complete
distribution of first passage times (FPTs). Our results demonstrate that the
typical FPT significantly differs from the MFPT, which
corresponds to the long time behaviour of the FPT distribution. Conversely, the
short time behaviour is shown to correspond to trajectories connecting directly
from the initial value to the target. Remarkably, we reveal a previously overlooked
third characteristic time scale of the first passage dynamics mirroring brief
excursion away from the target.

\end{abstract}


\section{Introduction}

How fast does the amplitude or position of a random process reach a given
threshold value (target) for the first time? This so-called first-passage
time (FPT) \cite{Sid,Ralf} is central to the description of the kinetics in
a large variety of systems across many disciplines, including diffusion controlled
chemical reactions \cite{Smol}, signalling cascades in biological cells
\cite{Alberts,otto}, transport in disordered media \cite{bAv} including the
breakthrough dynamics in hydrological aquifers \cite{brian}, the location of
food by foraging bacteria and animals \cite{Berg,Bell} up to the global spreading
of diseases \cite{Lloyd,dirkle} or stock market dynamics \cite{mantegna}. In the
following we discuss the FPT problem in the language of the diffusion of a physical
particle in position space.

Contrasting their diverse phenomenology, the kinetics in stochastic systems such as
the above can often
be rephrased in terms of the simplest---but extensively studied---random walk. In
unbounded space the FPT statistics of the random walk---or in fact its diffusion
limit---are heavy-tailed, giving rise to a diverging mean FPT (MFPT) \cite{Sid}.
Heavy tails are in fact common when it comes to persistence properties of infinite
systems \cite{Bray}. Conversely, a finite system size suppresses the heavy tails,
effecting an exponential long time statistic and thus a finite MFPT, which becomes
a function of the system size and dimensionality \cite{Sid,Olivier}. 

Generically we distinguish two universality classes of the statistic of the global
FPT (GFPT)---the FPT averaged over all initial positions inside the domain of
interest---for a variety of dynamics in translation invariant media, depending on
the nature of how the the surrounding space is explored \cite{Olivier,oNChem,oNat}:
in
the case of non-compact exploration leaving larger regions of the domain unexplored
such as in diffusion in three spatial dimensions, the initial separation between
the walker and its target does not play a dominant role \cite{oNChem}. The situation
is reversed in the case of compact exploration of space such as for the diffusion
on fractal geometries. Now the initial separation dominates and leads to so-called
geometry-controlled kinetics \cite{oNChem}. Note that for the statistic of the GFPT
the non-trivial dependence of the FPT statistics on the initial position is
effectively integrated out.

Many studies of FPT kinetics concentrate on the determination of the MFPT or the
GFPT which often are useful to determine the rough time scale of the underlying
process. However, even for Brownian motion the distribution of FPTs shows a very
rich phenomenology and---depending on the location of the target---may exhibit
highly non-uniform FPT kinetics \cite{Carlos1,Carlos2,Carlos3}. Under certain
conditions, any two independent first passage trajectories are most likely to be
significantly different. In such cases the MFPT---albeit finite---is not a
precise parameter to describe the FPT statistic \cite{Carlos1,Carlos2,Carlos3,
GlebRed}. Quite generally, the first passage statistic of Brownian motion has the
generic asymptotic behaviour \cite{Carlos2}
\numparts
\begin{eqnarray}
\label{generic1}
\lim_{t\to0}\wp(t)\simeq t^{-(1+\mu)}\exp(-a/t),\\
\lim_{t\to\infty}\wp(t)\simeq\exp(-bt),
\label{generic2}
\end{eqnarray}  
\endnumparts
where $\mu$ is the so-called persistence exponent \cite{Bray} and $a$ and $b$ are
dimension and geometry specific parameters. Eq.~(\ref{generic1}) encodes the fact
that it takes a finite minimum time to reach the target followed by a power law
decay of FPTs on a time scale on which the searcher does not yet feel the presence
of the boundary. In addition, Eq.~(\ref{generic2}) states that the searcher will
eventually find the target in a finite system of linear dimension $R$ on a time
scale up to $b\simeq R^2f_d(r_a)\gg a$, where $f_d(x)$ is a dimension dependent
function, which diverges as the target radius $r_a$ goes to zero in two and three
dimensions \cite{Carlos2}.  

The above results hold for translation invariant systems. Yet, numerous real systems
such as biological cells are spatially heterogeneous and therefore display
fundamentally different dynamics \cite{het1,het2,het3,het4}. Various aspects of
diffusion in heterogeneous media have already been addressed \cite{Thet1,Thet2,
Thet3,Thet4,hetero1,hetero2,hetero3} but the r{\^o}le of spatial heterogeneity in
the FPT statistics beyond the MFPT \cite{Het_AG} remains elusive. Moreover, the
results in Ref.~\cite{Het_AG} suggest that the MFPT in (hyper)spherically symmetric
domains is apparently independent of indirect trajectories, these are those that
interact with the confining boundary, in contrast to direct trajectories, that
head swiftly to the target.

Here we present exact results for the full FPT statistics in a simple heterogeneous
model system. Based on a rigorous asymptotic analysis of the FPT statistics of
Brownian motion in a confined spherically symmetric domain with a piece wise
constant diffusion coefficient, see Fig.~\ref{schm} and Ref.~\cite{Het_AG}, we
here demonstrate the emergence of a new time scale in the FPT dynamics, which
is controlled by the spatial heterogeneity. More precisely, we prove that the
intermediate time power law asymptotics in Eq.~(\ref{generic1}) breaks down in
a sufficiently heterogeneous medium. For such \emph{heterogeneity-controlled
kinetics\/} we derive exact asymptotic results for the short, intermediate, and
long time FPT statistics for an arbitrary degree of heterogeneity. We also quantify
the most likely (typical) FPT and the width of the FPT distribution. We demonstrate
that the MFPT is dominated by long and unlikely indirect trajectories, while the
overall relative contribution to the MFPT of the latter remains coupled to the
most likely, direct trajectories. Finally, we discuss the implications of our
results for more general systems.

\section{Results}

\subsection{System setup and general result}

We consider a spherically symmetric and potential free system with a perfectly
absorbing central target of finite radius $r_a$ and a perfectly reflecting boundary
at radius $R>r_a$ \cite{Het_AG}. The system is in contact with a heat bath at
constant and uniform temperature $T$. The particle experiences a space dependent
friction $\Gamma(r)$ originating from spatial variations in the long range
hydrodynamic coupling to the motion of the medium \cite{Oppenheim}. We focus on the
high friction limit corresponding to overdamped motion and assume that the particle
diffuses with the isotropic position dependent diffusion coefficient $D(r)=2k_BT
\times\Gamma(r)$. Simultaneously, the diffusing particle experiences the fluctuation
induced thermal drift $F(r)\sim k_BT\nabla\times\Gamma(r)$ ensuring thermodynamic
consistency in the sense that $D(r)$ has a purely stochastic origin and does not
reflect any heterogeneity in the entropic potential of mean force \cite{Lubensky}.
More precisely, in the absence of the target the system relaxes to the correct
Boltzmann-Gibbs equilibrium---a spatially uniform probability. The thermodynamically
consistent theory of diffusion in inhomogeneous media \cite{Lubensky} corresponds to
the so-called kinetic interpretation of the underlying multiplicative-noise Langevin
equation; see, for instance, Refs.~\cite{kinetic,kinetic2}.

\begin{figure}
\begin{center}
\includegraphics[width=12cm]{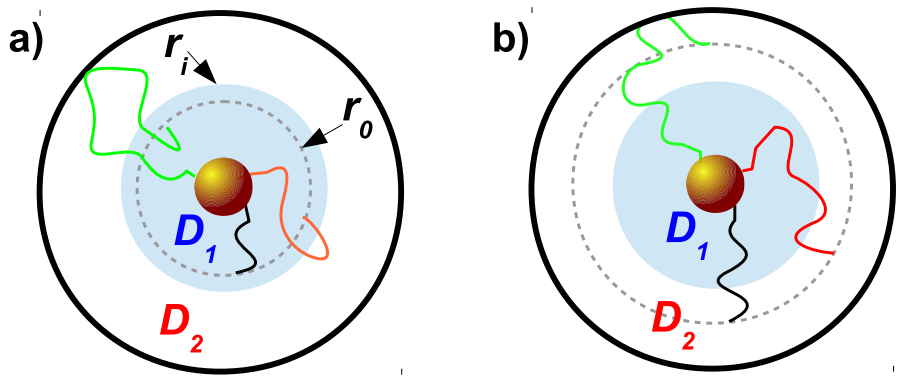}
\end{center}
\caption{Schematic of the model system with absorbing, finite central target of
radius $r_a$.
The concentric shell of radius $r_i$ separates regions of different diffusivities
$D_1$ and $D_2$. The initial radius of the particle (dashed line) is $r_0$, and
the outer shell at radius $R$ is reflective. The black and red lines denote direct
trajectories and the green line denotes an indirect trajectory, see text for
details.}
\label{schm}
\end{figure}

We are interested in the evolution of the probability density function
$P(r,t|r_0)$ in dependence on the particle radius $r$ at time $t$ after starting
from the initial radius $r_0$ at $t=0$. Due to the symmetry of the system the
angular co-ordinate is not of interest, and we average over the space angle.
The diffusion equation governing the radial probability density function $P(r,t|
r_0)$ is then given by
\begin{eqnarray}
\frac{\partial}{\partial t}P(r,t|r_0)=\frac{1}{r^2}\frac{\partial}{\partial r}
r^2D(r)\frac{\partial}{\partial r}P(r,t|r_0).
\label{diffeq}
\end{eqnarray}
In our analysis we consider the particular case of a piece wise constant diffusion
coefficient of magnitude $D(r)=D_1$ for $r\le r_i$ and $D(r)=D_2$ otherwise---see
also Fig.~\ref{schm}. Physically, this form should be viewed as an ideal limit of
a two phase system with a sharp interface. A similar limit of an infinitely sharp
interface can be taken for the Langevin equation of the process, as well.

The exact solution for the Laplace transform $\widetilde{P}(r,s|r_0)$ was
derived in Ref.~\cite{Het_AG}. From this result the Laplace transform of the FPT
density is obtained from the corresponding probability flux into the target,
$\widetilde{\wp}(s)=4\pi D_1r_a^2\times[\partial\widetilde{P}(r,s|r_0)/\partial r
]_{r=r_a}$. We introduce dimensionless variables $x=r/R$ for the particle position,
$x_a=r_a/R$ for the target radius, $x_i=r_i/R$ for the interface radius, and $x_0=
r_0/R$ for the initial particle position, as well as express time in units of
$R^2/\overline{D}$, where $\overline{D}$ is the spatial average of $D(r)$. The exact
result then reads
\numparts
\begin{eqnarray}
\nonumber
\hspace*{-1.2cm}
\widetilde{\wp}(s)&=&\frac{\sqrt{x_a/x_0}}{\displaystyle\frac{\mathcal{D}_{-1/2}(S_1
x_a,S_1x_i)}{\mathcal{C}_{-3/2}(S_2x_i,S_2)}+\frac{1}{\sqrt{\varphi}}\frac{
\mathcal{C}_{-1/2}(S_1x_a,S_1x_i)}{\mathcal{D}_{-1/2}(S_2x_i,S_2)}}\\
&\times&\left\{\begin{array}{ll}\displaystyle{\frac{\mathcal{D}_{-1/2}(S_1x_0,S_1
x_i)}{\mathcal{C}_{-3/2}(S_2x_i,S_2)}+\frac{1}{\sqrt{\varphi}}\frac{\mathcal{C}_{
-1/2}(S_1x_0,S_1x_i)}{\mathcal{D}_{-1/2}(S_2x_i,S_2)}}, & x_a< x_0\le x_i\\
\displaystyle{\frac{\mathcal{D}_{-1/2}(S_2x_0,S_2)}{x_iS_1\mathcal{D}_{-1/2}(S_2x_i
,S_2)\mathcal{C}_{-3/2}(S_2x_i,S_2)}}, & x_i <x_0\le 1\end{array}\right.,
\label{flux}
\end{eqnarray}
where we introduced the ratio $\varphi=D_1/D_2$ of the diffusivities along with the
abbreviation $S_{1,2}=\sqrt{s/D_{1,2}}$ and the auxiliary functions
\begin{eqnarray}
\nonumber
\mathcal{D}_{\nu}(z_1,z_2)&=&I_{\nu}(z_1)K_{\nu-1}(z_2)+K_{\nu}(z_1)I_{\nu-1}(z_2),\\
\mathcal{C}_{\nu}(z_1,z_2)&=&I_{\nu}(z_1)K_{\nu}(z_2)-I_{\nu}(z_2)K_{\nu}(z_1),
\label{Aux2}
\end{eqnarray}
\endnumparts
where $I_{\nu}$ and $K_{\nu}$ denote the modified Bessel functions of the first and
second kind, respectively \cite{Abramowitz}.

In order to allow for a meaningful comparison of the FPT kinetics at various degrees
$\varphi$ of heterogeneity we introduce a constraint on the conservation of the
spatially averaged diffusion coefficient, $\overline{D}=3(R^2-r_a^3)^{-1}\int_{r_a}
^Rr^2D(r)dr=const$, for a detailed discussion of this choice see Ref.~\cite{Het_AG}.
In addition, in the general case the absolute value of $D_1$ only sets the time
scale of the problem, whereas $\varphi$ gives rise to the qualitative changes in the
FPT kinetics in our heterogeneous system. With the introduced average diffusivity
constraint both diffusivities are fully determined by $\varphi$ and $x_i$, that is,
$Q^{-1}\equiv D_1/\overline{D}=\varphi/([\varphi-1]\chi(x_i)+1)$. Note, however,
that the constraint does not introduce any additional information which would affect
the qualitative picture of our results. The general case for arbitrary $D_1$ and
$D_2$ is recovered trivially by treating $Q$ as an independent parameter or by
replacing $1/Q\to D_1$ and $1/(\varphi Q)\to D_2$. 

Eq.~(\ref{flux}) is the starting point of our asymptotic analysis. In addition, in
order to validate the analytical results we numerically invert $\widetilde{\wp}(s)$
using the fixed Talbot method \cite{L_Tal}.

\subsection{Short time asymptotic}

Starting from the central result (\ref{flux}) the short time behaviour of the
FPT distribution $\wp(t)$ is obtained from the asymptotic behaviour of the
respective modified Bessel functions for large argument (large Laplace
variable) \cite{Abramowitz}. First it can be shown for $|z_{1,2}|\gg 1$ that
\begin{equation}
\mathcal{D}_{\nu}(z_1,z_2)\sim\frac{\cosh\left(z_1-z_2\right)}{\sqrt{z_1z_2}},
\,\,\,
\mathcal{C}_{\nu}(z_1,z_2)\sim\frac{\sinh\left(z_1-z_2\right)}{\sqrt{z_1z_2}},
\label{Aux3}
\end{equation}
which is
valid for $|\mathrm{arg}(z_{1,2})|<\pi/2$ and $|\mathrm{arg}(z_{1,2})|<3\pi/2$,
respectively. Combined with Eq.~(\ref{flux}) we obtain a limiting expression for
$\wp(t)$ which can be inverted exactly---by performing the contour integral
within the domain of validity of Eq.~(\ref{Aux3})---leading us to
the L{\'e}vy-Smirnov density
\numparts
\begin{equation}
\wp(t)\sim\frac{x_a}{x_0}\frac{\psi(x_a,x_i,x_0,\varphi)}{\theta(\varphi)}\sqrt{\frac{Q}{\pi t^3}}\exp\left(-\frac{Q\psi(x_a,x_i,x_0,\varphi)^2}{4t}\right)
\label{short}
\end{equation}
where we introduced 
\begin{equation}
\label{psi}
\psi(x_a,x_i,x_0)=\left\{\begin{array}{ll}x_0-x_a , & x_a< x_0\le x_i\\
x_i-x_a+\sqrt{\varphi}(x_0-x_i), & x_i <x_0\le 1\end{array}\right..
\end{equation}
\endnumparts
We take $\theta(\varphi)=2$ if $x_0\le x_i$ and $\theta(\varphi)=1+1/\sqrt{\varphi}$
otherwise. It is easy to see that all $x_i$-dependent terms vanish for a homogeneous
systems with $\varphi=1$. Note that Eq.~(\ref{short}) obeys the generic behaviour
given in Eq.~(\ref{generic1}) with a persistence exponent $\mu=1/2$ and $a=Q\psi(
x_a,x_i,x_0)/4$. Moreover, Eqs.~(\ref{short}) and (\ref{psi}) have an intuitive
physical meaning: they show that the earliest FPTs will be observed on a time
scale on which the particle diffuses over a distance corresponding to the initial
separation to the target, with the respective diffusion coefficients. Moreover, we
note that the FPT density to a \emph{finite} radius $x_a$ for three dimensional
radial Brownian motion in our heterogeneous system obeys the $t^{-3/2}$ scaling
Sparre Andersen theorem which needs to generally hold for one dimensional Markov
processes with symmetric jumps \cite{SA1,SA2}. Most importantly, Eq.~(\ref{psi})
is \emph{independent\/} of $D_2$ for $x_0<x_i$. In other words, the first passage
behaviour of particles released inside the interface radius $x_i$ is dominated by
trajectories, which head straight for the target and do not venture into the outer
part of the system. These are the \emph{direct trajectories\/} introduced in
Ref.~\cite{Het_AG}, see also below.

\begin{figure}
\begin{center}
\includegraphics[width=16cm]{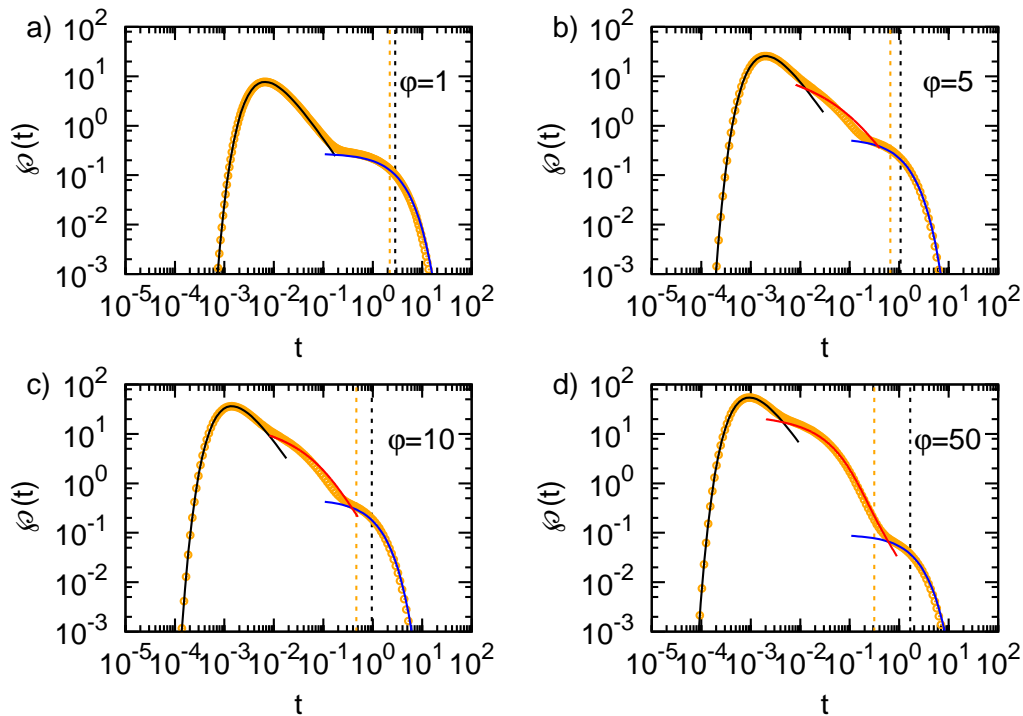}
\end{center}
\caption{FPT densities for various degrees of heterogeneity $\varphi$ and target
radius $x_a=0.1$, interface radius $x_i=0.5$, and initial radius $x_0=0.3$. In
this Figure the particle starts inside the inner region. The
symbols denote the results of the numerical inversion of Eq.~(\ref{flux}). The
black lines correspond to the short time limit (\ref{short}), the blue lines
denote the long time asymptotics given by  Eq.~(\ref{long}). The red lines
correspond to the intermediate time asymptotics in Eq.~(\ref{interm1}). The
dashed vertical lines denote the corresponding MFPT from $x_0$ (orange) and
from the outer boundary at $x=1$ (black), respectively.}
\label{fpt1}
\end{figure}

\begin{figure}
\begin{center}
\includegraphics[width=16cm]{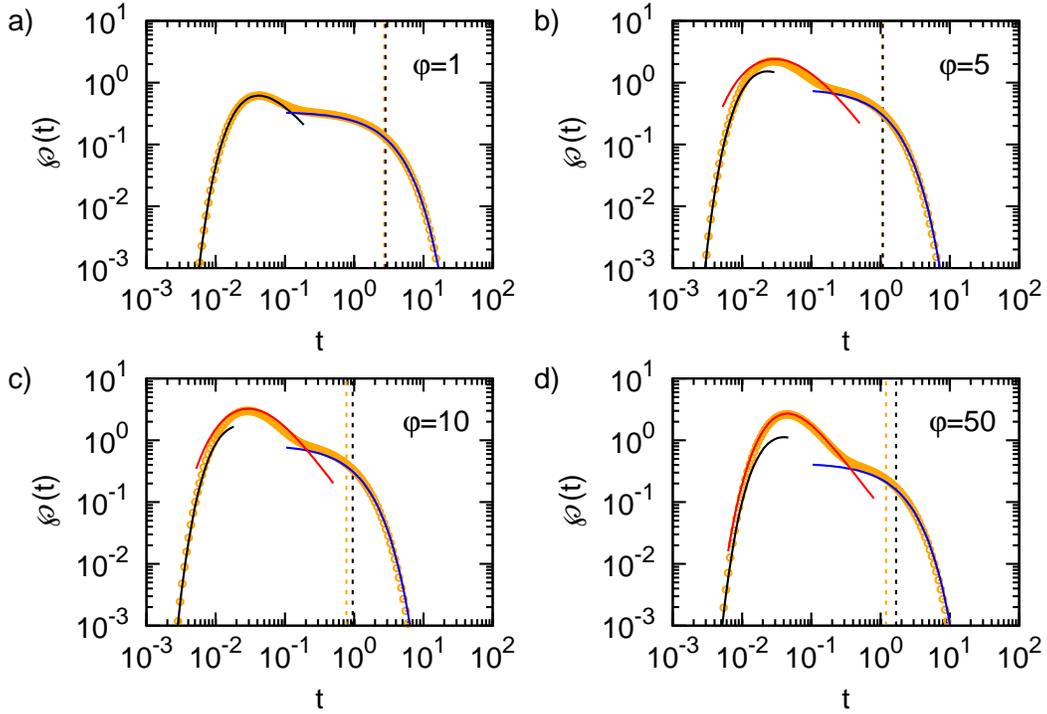}
\end{center}
\caption{FPT densities for various degrees $\varphi$ of heterogeneity and parameters
$x_a=0.1$, $x_i=0.4$ and $x_0=0.6$. In this Figure the particle starts in the outer
region of the system. The symbols denote the results from numerical inversion of
Eq.~(\ref{flux}). The black lines correspond to the short time limit (\ref{short})
and the blue line denotes the long time asymptotics (\ref{long}). The red line
represents the intermediate time asymptotics (\ref{interm2}). The dashed vertical
The dashed vertical line denotes the corresponding MFPT from $x_0$ (orange) and
from the outer boundary at $x=1$ (black), respectively.}
\label{fpt2}
\end{figure}

The FPT densities for various degrees of heterogeneity $\varphi$ are shown in
Figs.~\ref{fpt1} and \ref{fpt2}, corresponding to initial positions $x_0$ in the
inner and outer regions, respectively. Note the excellent agreement between the
exact numerical result for $\wp(t)$ and the short-time asymptotics in
Eq.~(\ref{short}) as seen from comparison of the symbols with the black lines. 

\subsection{Long time asymptotic}

The long time asymptotic behaviour of the FPT distribution $\wp(t)$ is obtained
from Eq.~(\ref{flux}) by using the expansions of the respective modified Bessel
functions for small argument (small Laplace variable $s$) \cite{Abramowitz}.
Here we strictly note
that corrections to leading order terms in $I_{\nu}$ and $K_{\nu}$ must be retained
in order to obtain the correct behaviour of $\mathcal{D}_{\nu}$ and $\mathcal{C}_{
\nu}$ for small $s$. We find that
\begin{eqnarray}
\nonumber
\mathcal{D}_{-1/2}(\sqrt{s}y_1,\sqrt{s}y_2)&\sim&\frac{1}{y_2}\sqrt{\frac{y_1}{
y_2s}}+\frac{(y_2-y_1)^2(y_1+2y_2)\sqrt{s}}{6y_2\sqrt{y_1y_2}}\\
\nonumber
\mathcal{C}_{-1/2}(\sqrt{s}y_1,\sqrt{s}y_2)&\sim&\frac{y_1-y_2}{\sqrt{y_1y_2}}+
\frac{(y_1-y_2)^3}{6\sqrt{y_1y_2}}s\\
\mathcal{C}_{-3/2}(\sqrt{s}y_1,\sqrt{s}y_2)&\sim&\frac{y_1^3-y_2^3}{3(y_1y_2)^{
3/2}}+\frac{(y_1-y_2)^3(y_1^2+3y_1y_2+y_2^2)}{30(y_1y_2)^{3/2}}s.
\label{Aux4}
\end{eqnarray}
Combining Eqs.~(\ref{Aux4}) with Eq.~(\ref{flux}) we invert the Laplace transform
exactly, yielding the exponential density
\begin{equation}
\wp(t)\simeq \frac{\langle t_{x_a}(x_0)\rangle}{\langle t_{x_a}(1)\rangle^2}\exp
\left(-\frac{t}{\langle t_{x_a}(1)\rangle}\right),
\label{long}
\end{equation}
where the symbol $\langle t_{x_a}(x)\rangle$ denotes the MFPT to $x_a$ if
starting from $x$. Its result is given by
\begin{eqnarray}
\label{psiL}
\langle t_{x_a} (x)\rangle=Q\left\{\begin{array}{ll}\langle t^0_{x_a}(x)\rangle,
& x_a<x\le x_i\\\langle t^0_{x_a}(x_i)\rangle+\varphi\langle t^0_{x_i}(x)\rangle,
& x_i<x\le1\end{array}\right.,
\end{eqnarray}
where $\langle t^0_{x_a}(x_0)\rangle$ denotes the MFPT from $x_0$ to $x_a$ in a
\emph{homogeneous\/} sphere with unit radius and with unit diffusion coefficient,
\begin{equation} 
\label{MFPT}
\langle t^0_{x_a} (x)\rangle=\frac{1}{3}\left(\frac{1}{x_a}-\frac{1}{x}+\frac{1}{2}
[x_a^2-x^2]\right).
\end{equation} 
As before, all $x_i$-dependent terms vanish for a homogeneous system.

Eq.~(\ref{long}) has the generic form of the exponential long time tail
(\ref{generic2}) of the FPT distribution in a finite system. From Eq.~(\ref{long})
we identify the inverse of the characteristic time as $b=\langle t_{x_a}(1)\rangle
^{-1}$.
Moreover, Eq.~(\ref{long}) has an intuitive meaning: it demonstrates that first
FPTs are exponentially unlikely beyond a time scale corresponding to the
MFPT to arrive at the target from the external boundary. In turn, the long time
exponential region evidently corresponds to trajectories, which are reflected from
the external boundary. Therefore, for $x_0=1$ there we can no longer distinguish
between direct and indirect trajectories. Note the excellent
agreement between the exact numerical result for $\wp(t)$ and the long time
asymptotics (\ref{long}) in Figs.~\ref{fpt1} and \ref{fpt2} as shown by the blue
lines.

Moreover, we emphasise another observation. For a homogeneous system with $\varphi
=1$ the short and long time asymptotics together fully describe the FPT density.
Put differently the overlap region between the regimes is extremely narrow. This
holds true in general up to some critical heterogeneity $\varphi^{\star}$, which
will be specified in the following section. Beyond this value $\varphi^{\star}$ a new
time scale emerges, as seen in Fig.~\ref{fpt1}b-d and Fig.~\ref{fpt2}b-d) which is
not captured by the the short and long time asymptotics and does \emph{not\/}
correspond to an overlap regime, as we now explain.

\subsection{Emergence of a new time scale}

Here we focus on the regime $\varphi>1$ when the inner region has the higher
diffusivity. This is the scenario which we would naively expect to enhance the FPT
kinetics. The opposite case $\varphi<1$ is physically less interesting but can be
obtained
analogously to the steps presented below. Considering the two different types of
argument in Eq.~(\ref{flux}), $\sqrt{sQ}x_k$ and $\sqrt{s\varphi Q}x_l$ it becomes
obvious that an additional separation of time scales occurs in the limit $Qx_k^2\ll
s^{-1}\ll Qx_l^2$, where $k,l$ stand for the different indices used in our model.
In other words, there exists a time scale
separation between direct trajectories corresponding to the short time asymptotic
(\ref{short}) and reflected trajectories accounted for by the long time asymptotic
(\ref{long}).

This new time scale corresponds to trajectories, which are much longer than the
direct ones yet much shorter than the reflected ones. In such trajectories
the particle ventures into the outward direction away from the target with respect
to its initial
position. However, this excursion is much shorter than the average time needed to
reach the interface. The result are terms of mixed order in $s$ having the form
\begin{equation}
1+\frac{x_k}{x_l}(x_l-x_k)\sqrt{\frac{Qs}{\varphi}}+\frac{(x_l-x_k)^2(1+2x_l/x_k)
Qs}{6}
\label{iAux}
\end{equation}
with $x_l>x_k$. In this limit the second and third terms are comparable and both need
to be explicitly considered in the analysis. The exact intermediate time asymptotic
forms of $\wp(t)$ can be derived rigorously and read
\numparts
\begin{eqnarray}
\nonumber
\wp_<(t)&\sim&\mathcal{A}_1/\varphi^2\Bigg(\frac{\mathcal{A}_2}{\sqrt{\pi t}}+\sum
_{n=0}^{\infty}(-1)^n\left(\mathcal{A}_3\sqrt{t}/\mathcal{A}_4\right)^{2n+1}\\
&\times&\left[\mathcal{A}_5U\left(1+n,1/2,t/\mathcal{A}_4^2\right)-\mathcal{A}_6U
\left(1+n,3/2,t/\mathcal{A}_4^2\right)\right]\Bigg),
\label{interm1}
\end{eqnarray}
when $x_0\le x_i$ and 
\begin{eqnarray}
\nonumber
\wp_>(t)&\sim&\mathcal{B}_1\exp\left(-\mathcal{B}_2^2/t\right)\sum_{n=0}^{\infty}
\left(-\mathcal{B}_3t\right)^{n}\\
\nonumber
&\times&\Bigg[\left(\mathcal{B}_2/\sqrt{t}+\mathcal{B}_4\sqrt{t}\right)U\left(1+n,
3/2,\left[\mathcal{B}_2/\sqrt{t}+\mathcal{B}_4\sqrt{t}\right]^2\right)\\
&-&\mathcal{B}_4\sqrt{t} U\left(1+n,1/2,\left[\mathcal{B}_2/\sqrt{t}+\mathcal{B}_4
\sqrt{t}\right]^2\right)\Bigg],
\label{interm2}
\end{eqnarray}
\endnumparts
when $x_0>x_i$ as well as for $t>\mathcal{B}_2$. Here, $U(\alpha,\beta,z)$ denotes
Tricomi's confluent hypergeometric function \cite{Abramowitz}. The coefficients
$\mathcal{A}_1,\ldots,\mathcal{A}_6$ and $\mathcal{B}_1,\ldots,\mathcal{B}_4$
are given in Eqs.~(\ref{TheAs}) and (\ref{TheBs}) in the Appendix. Details of
the calculation will be reserved for a separate longer publication. Both forms,
Eq.~(\ref{interm1}) and Eq.~(\ref{interm2}) hold for $\varphi>\varphi^{\star}$ with
\begin{equation}
\varphi^{\star}=\frac{3x_i}{2x_a}\times\left(2+\frac{3x_a}{2x_i}\right)^{-1},
\label{fic}
\end{equation}
independently of $x_0$. The transition as $\varphi$ crosses the critical value 
$\varphi^{\star}$ is discontinuous, the factor involved changing from $\sqrt{1-2
\varphi x_a(2+x_a/x_i)/3x_i}$ to $\sqrt{2\varphi x_a(2+x_a/x_i)/3x_i-1}$, and the
functional dependence on $t$ changes concurrently as well. Below
$\varphi^{\star}$ the FPT distribution $\wp(t)$ is completely specified in terms of
Eq.~(\ref{short}) and Eq.~(\ref{long}). This is what we may call a \emph{threshold
heterogeneity}. Moreover, while Eq.~(\ref{interm1}) does not reduce to a simple form
as $\varphi\to\infty$, Eq.~(\ref{interm2}) reduces to a L{\'e}vy-Smirnov density,
\begin{equation}
\lim_{\varphi\to\infty}\wp_>(t)\simeq Kt^{-3/2}\exp\left(-\varphi Q\frac{(x_0-x_i)
^2}{4t}\right)
\label{a_interm2}
\end{equation}
with a computable prefactor $K$ different from the quantity $\psi(x_i,x_i,x_0,
\varphi)/\theta(\varphi)$ in Eq.~(\ref{short}). Other than that, Eq.~(\ref{interm2})
continuously interpolates between the two different L{\'e}vy-Smirnov densities
Eqs.~(\ref{short}) and (\ref{a_interm2}) as $\varphi$ increases. Intuitively, this
latter limit demonstrates the fact, that reaching the interface from $x_0>x_i$
becomes rate limiting for large $\varphi$. We also note that the infinite series in
Eqs.~(\ref{interm1}) and  (\ref{interm2}) converge fast, and in numerical evaluations
it suffices to consider the first 10--15 terms for any value of $\varphi$. The
intermediate asymptotic formulas are compared to the exact numerical result in
Figs.~\ref{fpt1} and \ref{fpt2}, as shown by the symbols and the red line.
We find a good agreement, which intuitively depends on the separation of time scales
and hence improves for large $\varphi$. Strikingly, for $x_0>x_i$ the intermediate
time regime includes the \emph{most likely\/} FPTs. Therefore, most likely
trajectories are indeed direct, as we anticipated already in \cite{Het_AG}. 

\begin{figure}
\begin{center}
\includegraphics[width=16cm]{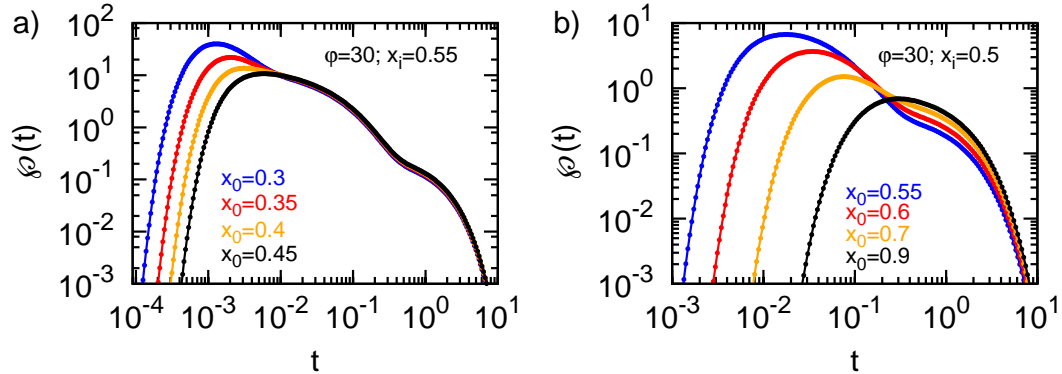}
\end{center}
\caption{FPT densities for target radius $x_a=0.1$ and various combinations of
the starting position $x_0$ and the interface radius $x_i$, as denoted in the
panels.}
\label{em}
\end{figure}

To gain more intuition on how exactly the time scale separation arises we plot the
exact numerical result for the FPT distribution $\wp(t)$ at $\varphi=30$ for
different initial positions, as depicted in Fig.~\ref{em}. Starting in the inner
region the scale separation emerges as we continuously move the initial position
towards the target. Conversely, if starting in the outer region a scale separation
emerges as we continuously move the initial position away from the reflecting
surface. The necessary and sufficient requirement for heterogeneity controlled
kinetics is therefore a large heterogeneity and a corresponding existence of two
length scales inwards and outwards from the initial position towards the closest
boundary or interface. 

\subsection{Mean first passage times are not typical}

We now quantify the most likely or typical FPT times, i.e., those which occur most
frequently. Already from Figs.~\ref{fpt1} and \ref{fpt2} it is apparent that there
is a large discrepancy in the likelihood of typical and mean FPTs, compare
the maximum of $\wp(t)$ with the dashed vertical line denoting the MFPT. In many
cases the most likely FPT is the more relevant quantity. For instance, consider a
certain species of bacteria, in which genetic regulation can be viewed as an FPT
problem \cite{otto}. When we compare the fitness of an individual bacteria in a
colony, those who are among the first to respond to an external challenge will be
of advantage. Similarly, those predators that first discover a prey have a larger
chance of survival. 

\begin{figure}
\begin{center}
\includegraphics[width=16cm]{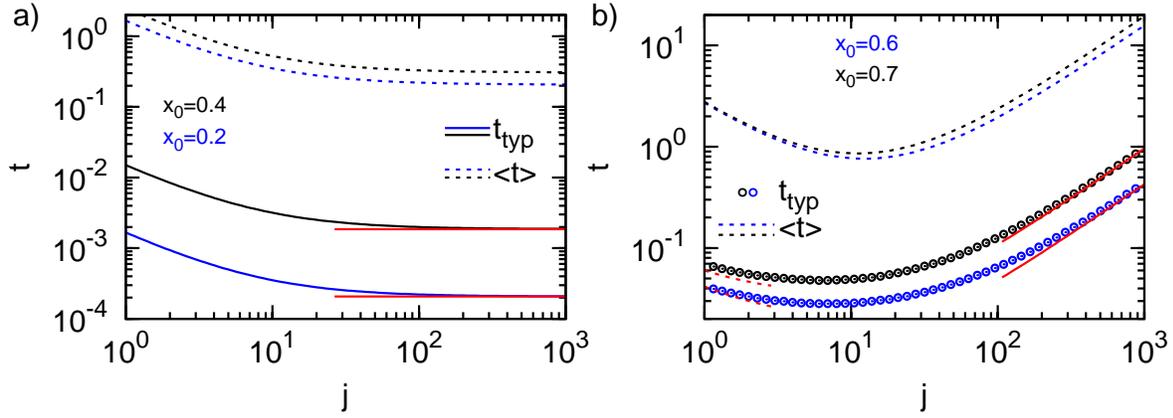}
\end{center}
\caption{Comparison of the MFPT and the most likely FPT for two different initial
conditions and target radius $x_a=0.1$: a) $x_0<x_i$ for $x_i=0.5$ and b) $x_0>x_i$
for $x_i=0.4$.
The full and dashed black and blue lines denote the MFPT and the most likely FPT,
respectively. The red lines correspond to Eqs.~(\ref{typ2}).}
\label{typ_m}
\end{figure}

The MFPT was defined in the previous section, while the typical FPT is corresponds
to the extremum $d\wp(t)/dt=0$. In the case $x_0\le x_i$ we find that
\begin{equation}
t_{\mathrm{typ}}=\frac{Q(x_0-x_a)^2}{6},
\label{typ1}
\end{equation}  
whereas in the case $x_0>x_i$ we obtain closed form expressions only in the limits
$\varphi\simeq1$ and $\varphi\gg1$, these being
\numparts
\begin{eqnarray}
\lim_{\varphi\to1^+}t_{\mathrm{typ}}&=&\frac{Q(x_0-x_a+\sqrt{\varphi}[x_0-x_i])
^2}{6}\\
\lim_{\varphi\to\infty}t_{\mathrm{typ}}&=&\frac{Q(x_0-x_i)^2}{6}.
\label{typ2}
\end{eqnarray}
\endnumparts
For general values of $\varphi$ the typical FPTs are computed numerically.

The typical and mean FPTs as a function of $\varphi$ are compared in
Fig.~\ref{typ_m}. We observe that $t_{\mathrm{typ}}$ is 2--3 orders of magnitude
smaller than the corresponding MFPT. Moreover, the amplitude of the FPT distribution
at the positions of $t_{\mathrm{typ}}$ and of the MFPT differs by around one order
of magnitude. The MFPT hence is an imprecise measure for the FPT kinetics of Brownian
motion in both homogeneous and heterogeneous media. Typical trajectories are hence
direct in the sense that they do not reach the external boundary. In fact, provided
with the results for the three time scales of the current FPT problem obtained in
the previous sections we are now in the position to make a more precise statement:
\emph{Typical first FPTs are strictly shorter than the average time needed to
arrive at the target via reflection from the interface.} Note that this statement
does not contradict the case of starting at the reflecting surface, as the first
reflection would demand a return to the initial position.

Conversely, it is obvious that while the long time FPT behaviour \emph{per se\/} is
completely independent of the initial particle position, the contribution of this
regime to the MFPT depends on the relative time span of the regime in comparison to
the short and intermediate time regimes. In other words, the long time regime always
has a strictly additive contribution, whose relative magnitude does depend on $x_0$
through the lower limit of integration. To see this we can approximately split the
MFPT into a long time contribution $\wp(t)_l$ and the remainder $\wp(t)_r$, where
the respective parts hold for $t\le t^{\ast}$ and $t> t^{\ast}$. Then from
Eqs.~(\ref{long}) we find that
\begin{equation}
\label{split}
\langle t_{x_a}(x)\rangle=\int_0^{\infty}t\wp(t)dt\sim\int_0^{t^{\ast}}t\wp_r(t)dt
+\int_{t^{\ast}}^{\infty}t\wp_l(t)dt,
\end{equation}
where the second term depends on $x_0$ solely through $t^{\ast}$. Using
Eq.~(\ref{long}) and performing the integral we find that the long time
contribution to the MFPT---the second term in Eq.~(\ref{split})---is always equal to
\begin{equation}
\label{c_long}
\langle t_{x_a}(x_0)\rangle_l=\langle t_{x_a}(x_0)\rangle\exp\left(-\frac{t^{\ast}
}{\langle t_{x_a}(1)\rangle}\right)\times\left(1+\frac{t^{\ast}}{\langle t_{x_a}(1)
\rangle}\right).
\end{equation}
It is straightforward to check that the long time exponential tail dominates the
MFPT. This finding is \emph{a priori\/} puzzling as it appears to be incompatible
with the additivity of the MFPT in Eq.~(\ref{MFPT}). This can be resolved from
cognisance of the fact that the second integrand is always the exact MFPT scaled
by a unit exponential $\langle t_{x_a}(x_0)\rangle\exp(-y)$, and the integration
is over $y$ from $t^{\ast}/\langle t_{x_a}(1)\rangle$ to $\infty$. It should be
noted that for most physically realistic situations we have $\langle t_{x_a}(x_0)
\rangle_l\simeq\langle t_{x_a}(x_0)\rangle$. We are hence in the position to make
the quite powerful statement, refining the results in Ref.~\cite{Het_AG}:
\emph{Despite being dominated by the rare long time, indirect trajectories the MFPT
is in fact completely specified by the statistics of direct trajectories. Namely,
the fraction and duration of direct trajectories rescales the otherwise invariant
contribution of indirect ones. In other words, the value of the integral is
essentially constant but its overall statistical weight is set by the direct
trajectories. This also explains the result (\ref{MFPT}).} Moreover, the
non-monotonic behaviour with respect to $\varphi$ for $x_0>x_i$ and the
corresponding existence of a minimum can be understood intuitively in terms of the
balance between the rate to arrive from $x_0$ to the interface and the rate to
arrive from $x_i$ to the target, compare also Ref.~\cite{Het_AG}.

We finally quantify the width $\Delta$ of the FPT distribution $\wp(t)$. Since we
know that very short and very long trajectories are exponentially unlikely, we may
use these cutoff times obtain
\begin{equation}
\label{width} 
\Delta=\langle t_{x_a}(1)\rangle-\frac{3}{2}t_{\mathrm{typ}},
\end{equation}
such that the width depends on the initial position solely through $t_{\mathrm{
typ}}$. Moreover, we find that
\begin{eqnarray}
\label{Lwidth} 
\lim_{\varphi\to\infty}\Delta\sim\varphi Q\times\left\{\begin{array}{ll}\langle
t^0_{x_i}(1)\rangle, & x_a<x_0\le x_i\\[0.2cm]
\displaystyle\left(1/x_i+3x_0x_i/2-[x_i^2+3x_0^2]/4\right)/3, & x_i<x_0\le1
\end{array}\right.,
\end{eqnarray} 
demonstrating that the width increases as $\varphi$ grows and effects a progressive
retardation of the dynamics in the outer region.

\section{Discussion}

We analysed a simple prototype model for Brownian motion in heterogeneous
environments in terms of a spherically symmetric geometry with two concentric
regions of different diffusivity. We obtained rigorous, asymptotically exact
results for the FPT distribution and identified a short time L{\'e}vy-Smirnov
as well as a long time exponential behaviour. Moreover, we demonstrated the
existence of a hitherto overlooked intermediate time scale. All three time scales
were interpreted in terms of direct and indirect trajectories. The distinction
between the latter also effects the major discrepancy between the mean and the
most likely FPTs. Cognisance of this difference is important in many systems.

We first focus on the implications of our findings for Brownian motion in
spherically symmetric homogeneous media. Our results suggest that the prevailing
paradigm of first passage kinetics for Brownian motion \cite{Sid,Olivier,oNChem}
becomes even richer. First, as highlighted already in previous works \cite{Carlos1,
Carlos2,Carlos3} the MFPT is often a rough measure of the first passage kinetics:
the FPT distribution is typically positively skewed and therefore asymmetric.
According to the results in Ref.~\cite{Carlos1,Carlos2,Carlos3} any two arbitrary
trajectories are often more likely to be very different than similar. From our
results reported here this result is substantiated in the sense that according to
our interpretation an arbitrary trajectory will typically be direct and will not
interact with the reflecting boundary. For homogeneous Brownian motion this holds
strictly for all starting positions satisfying $\wp(t_{\mathrm{typ}})Q\langle t_{
x_a}^0(1)\rangle^2>\langle t^0_{x_a}(x_0)\rangle$. This statement is particularly
important for single molecule observations, where a finite number of trajectories
will more likely reveal the typical and not the mean behaviour. We have shown that
there is a large discrepancy between the typical FPT and the MFPT. Moreover, an
upper bound for the FPT---if starting form an arbitrary position---is set by the
MFPT from the confining surface. In the latter case, the distinction between direct
and indirect trajectories obviously ceases to exist, giving rise to dominantly
exponential statistics. Conversely, the MFPT, while indeed dominated by the rare 
long time behaviour, is remarkably fully specified by the typical direct
trajectories. Hence, the most likely but less significant trajectories turn out to
determine the unlikely but dominant trajectories. This adds another surprising
feature to the first passage behaviour of Brownian motion, starting with the
L{\'e}vy arcsine laws for one dimensional free Brownian motion \cite{Levy,SatyaCS}.

We demonstrated that a sufficient heterogeneity in the diffusion coefficient gives
rise to an additional, intermediate time scale, on which trajectories contain a
short excursion towards the external surface. This excursion, however, is much
shorter than the typical time needed to diffuse across the entire domain. In this
\emph{heterogeneity-controlled\/} kinetic domain there thus exist three distinct
classes of trajectories: the direct ones, the indirect ones and those which
initially make a short indirect excursion and then go directly to the target, as
shown in Fig.~\ref{schm}. This latter class of trajectories in fact represents the
typical trajectories for starting positions in the outer region of our model system.

The results for our idealised two component model studied in the present work have
important consequences for an arbitrary spherically symmetric modulation of the
diffusivity. Namely, in such a case there might exist several distinct time scales
in the heterogeneity controlled kinetic regime. In fact, heterogeneity controlled
kinetics will generically be observed in the presence of sharp changes in the
local diffusivity $D(r)$. Such sharp modulations (dynamic surfaces) are indeed
present in cellular signalling processes when a particle starts in the cytoplasm
and searches for its target in the nucleus, where the diffusivity is different.
In particular, signalling particles
synthesised inside the cell as part of a particular signalling cascade
will inherently start away from the cellular membrane, and a
separation of scales in the FPT is therefore expected to exist. Moreover, as a
large discrepancy is expected to exist in the likelihood for observing typical FPTs
with respect to the MFPT, the MFPT is a particularly fairly poor measure for the
kinetics at low copy numbers of signalling molecules and for a finite number of
possible realisations.
Conversely, smooth modulations of $D(r)$ are not expected to change the
qualitative two time scale picture of FPT kinetics but will of course alter the
coefficients in the short and long time asymptotic regimes. We can extend the
discussion also to systems with off-centre targets under the condition that the
searching particle does not start too close to the external boundary. In this case
the long and intermediate time scale asymptotics would remain unchanged---as long
as there exist a separation of scales, of course---but the long time asymptotics
would be altered. Thus, our rigorous results for our idealised model system are
relevant for FPT kinetics in generic heterogeneous media.   

We also briefly comment on alternative forms of the diffusion equation
(\ref{diffeq}), the so called Ito and Stratonovich forms \cite{kinetic}. Here
we focus on physical stochastic dynamics satisfying the fluctuation-dissipation
theorem. In this view, the Ito and Stratonovich equations are nothing but the
corresponding Fokker-Planck equations with diffusion coefficient $D(r)$ in an
external potential $U(r)=-k_BT\ln D(r)$ and $U(r)=-(k_BT/2)\ln D(r)$, respectively.
The FPT kinetics in the presence of such effective external fields will therefore
be fundamentally different. We finally note that it will be of further interest 
to include intermittent active motion in the analysis \cite{active}.

\ack

AG acknowledges funding through an Alexander von Humboldt Fellowship and ARRS
programme P1-0002. RM acknowledges funding from the Academy of Finland (Suomen
Akatemia) within the Finland Distinguished Professor scheme.

\begin{appendix}

\section{Explicit expressions for the coefficients $\mathcal{A}_i$ and
$\mathcal{B}_i$}

The coefficients in Eq.~(\ref{interm1}) read
\begin{eqnarray}
\label{TheAs}
\mathcal{A}_1&=&\frac{9}{\varphi^2}\left[\Lambda_1(x_a,x_i,x_0)^2\Lambda_2(x_i,x_i
,x_a)^2\Lambda_2(x_i,x_i,x_0)\Lambda_2(x_i,x_a,x_i)\right]^{-1}\nonumber\\
\mathcal{A}_2&=&\frac{\varphi}{6}\Lambda_2(x_a,x_a,x_0)\left(1+[x_a+x_0]/x_i\right)
\nonumber\\
\mathcal{A}_3&=&\sqrt{2\varphi x_a(2+x_a/x_i)/3x_i-1}\nonumber\\
\mathcal{A}_4&=&\frac{3}{\varphi}(2+x_a/x_i)/\Lambda_2(x_a,x_a,x_i)\nonumber\\
\mathcal{A}_5&=&\frac{1}{4\mathcal{A}_3}\left(1-\mathcal{A}_3-\Lambda_1(x_0,x_i,
x_a)\left[\varphi\Lambda_2(x_0,x_i,x_a)/3\right.\right.\nonumber\\
&&-\{\mathcal{A}_3-1\}/\left.\left.\Lambda_1(x_a,x_i,x_0)\right]\right)\nonumber\\
\mathcal{A}_6&=&\frac{1}{4\mathcal{A}_3}\left(2-\Lambda_1(x_0,x_i,x_a)\left[2/
\Lambda_1(x_a,x_i,x_0)+\varphi\Lambda_2(x_0,x_i,x_a)/3\right]\right)
\end{eqnarray}
where we have introduced the auxiliary functions
\begin{equation*}
\Lambda_1(x,y,z)=(2+x/y)/(y-z)\qquad\Lambda_2(x,y,z)=x(z/y-1).
\end{equation*}
Conversely, the coefficients in Eq.~(\ref{interm2}) are
\begin{eqnarray}
\label{TheBs}
\mathcal{B}_1&=&\varphi\frac{x_i}{x_0}\Lambda_2(x_i,x_a,x_i)^{-2}\nonumber\\
\mathcal{B}_2&=&\frac{\sqrt{\varphi Q}}{2}(x_0-x_i)\nonumber\\
\mathcal{B}_3&=&\frac{6}{Q}\left[\Lambda_2(x_a,x_i,x_a)^2(1+2x_i/x_a)\right]^{-1}
\left(1-3(x_i/x_a)^2\{2\varphi(1+2x_i/x_a)\}^{-1}\right)\nonumber\\
\mathcal{B}_4&=&3\frac{x_i}{x_a}\left[\sqrt{Q\varphi}\Lambda_2(x_a,x_i,x_a)(1+2x_i
/x_a)\right]^{-1}.
\end{eqnarray}

\end{appendix}


\section*{References}

\begin{thebibliography}{99}

\bibitem{Sid} Redner S 2001 \emph{A Guide to First Passage Processes}
(Cambridge University Press, New York, USA). 

\bibitem{Ralf} Metzler R, Oshanin G and Redner S (Eds) 2014 \emph{First-Passage
Phenomena and Their Applications} (World Scientific, Singapore). 

\bibitem{Smol} von Smoluchowski M 1916 \emph{Phys. Z.} \textbf{17}, 557.

\bibitem{Alberts} Alberts B et al. 2002 \emph{Molecular Biology of the Cell}
(Garland, New York, NY).

\bibitem{otto} Pulkkinen O and Metzler R 2013 \emph{Phys. Rev. Lett.}
\textbf{110}, 198101.

\bibitem{bAv} ben-Avraham D and Havlin S 2000 \emph{Diffusion and
Reactions in Fractals and Disordered Systems} (Cambridge University Press,
Cambridge, UK). 

\bibitem{brian} Berkowitz B, Cortis A, Dentz M and Scher H 2006 \emph{Rev.
Geophysics} \textbf{44}, RG2003.

\bibitem{Berg} Berg HC 1993 \emph{Random Walks in Biology} (Princeton University
Press, Princeton). 

\bibitem{Bell} Bell WJ 1991 \emph{Searching Behaviour} (Chapman \& Hall, London). 

\bibitem{Lloyd} Lloyd AL and May RM 2001 \emph{Science} \textbf{292}, 1316 (2001).

\bibitem{dirkle} Hufnagel L, Brockmann D and Geisel T (2004) \emph{Proc. Natl.
Acad. Sci. USA} \textbf{101}, 15124.

\bibitem{mantegna} Mantegna RN and Stanley HE 2007 \emph{Introduction to
Econophysics: Correlations and Complexity in Finance} (Cambridge University
Press, Cambridge, UK).

\bibitem{Bray} Bray AJ, Majumdar SN and Schehr G 2013 
  \emph{Adv. Phys.} \textbf{62}, 325.

\bibitem{Olivier} B\'enichou O and Voituriez R 2014 \emph{Phys. Rep.} \textbf{539}, 225.

\bibitem{oNChem} B\'enichou O, Chevalier C, Klafter J, Meyer B and
  Voituriez R 2010 \emph{Nature Chem.}  \textbf{2}, 472.

\bibitem{oNat} Condamin S, B\'enichou O, Tejedor V, Voituriez R and
  Klafter J 2007 \emph{Nature} \textbf{450}, 77.

\bibitem{Carlos1} Mej\'ia-Monasterio C, Oshanin G and Schehr G 2011
  \emph{Phys. Rev. E} \textbf{84}, 035203.

\bibitem{Carlos2} Mej\'ia-Monasterio C, Oshanin G and Schehr G 2011
  \emph{J. Stat. Mech.} \textbf{85}, P06022.

\bibitem{Carlos3} Mattos TG, Mej\'ia-Monasterio C, Metzler R and
  Oshanin G 2012 
  \emph{Phys. Rev. E} \textbf{86}, 031143.

\bibitem{GlebRed} Oshanin G and  Redner S 2009
  \emph{Europhys. Lett.} \textbf{85}, 10008.


\bibitem{het1} K\"uhn T et al. 2011 \emph{PLoS ONE}
  \textbf{6}, e22962. 

\bibitem{het2} English BP et al. 2011 \emph{Proc. Natl. Acad. Sci. USA}
  \textbf{108}, E365. 

\bibitem{het3} Serg\'e A, Bertaux N, Rigenault H and Marguet D 2008 \emph{Nature Methods}
  \textbf{5}, 687.

\bibitem{het4} Cutler PJ et al. 2013 \emph{PLoS ONE}
  \textbf{8}, e64320. 

\bibitem{Thet1} Cherstvy AG, Chechkin AV and Metzler R 2013 \emph{New J. Phys.}
\textbf{15}, 083039.

\bibitem{Thet2} Cherstvy AG and Metzler R. 2014 \emph{Phys. Rev. E} \textbf{90}, 012134. 

\bibitem{Thet3} Massignan P et al. 2013 \emph{Phys. Rev. Lett.}
\textbf{15}, 083039. 
 
\bibitem{Thet4} Gu\'{e}rin T and Dean DS 2015 \emph{Phys. Rev. Lett.}
\textbf{115}, 020601.

\bibitem{hetero1} Godec A, Bauer M and Metzler R 2014 \emph{New J. Phys.} \textbf{16}, 092002.

\bibitem{hetero2} Gouze M, Melean Y, Le Borgne T,
  Dentz M and Carrera J 2008 \emph{Water. Resour. Res.} \textbf{44}, W11416. 

\bibitem{hetero3} Ukmar T, Gaber\v{s}\v{c}ek M, Merzel F and
  Godec A 2011 \emph{Phys. Chem. Chem. Phys.} \textbf{13}, 15311.

\bibitem{Het_AG} Godec A and Metzler R 2015 \emph{Phys Rev E} \textbf{91}, 052134.

\bibitem{Oppenheim} Tokuyama M and Oppenheim I 1995 \emph{Physica A} \textbf{216}, 85.

\bibitem{Lubensky} Lau AWC and Lubensky TC 2007 \emph{Phys. Rev. E}
  \textbf{76}, 011123.

\bibitem{kinetic} Klimontovich YL 1990 \emph{Physica A} \textbf{163}, 515 

\bibitem{kinetic2} H\"anggi P and Thomas H 1982 \emph{Phys. Rep.} \textbf{88}, 207. 

\bibitem{Abramowitz} Abramowitz M and Stegun IA 1970 \emph{Handbook of
Mathematical Functions} (Dover, New York, NY).

\bibitem{L_Tal} Abate J, Valk\'{o} PP 2004 \emph{Int. J. Numer. Meth. Engng.}
\textbf{60}, 979.

\bibitem{SA1} Sparre Andersen E 1953 \emph{Math. Scand.} \textbf{1}, 263. 

\bibitem{SA2} Sparre Andersen E 1954 \emph{Math. Scand.} \textbf{2}, 195.

\bibitem{Levy} L{\'e}vy P 1939 \emph{Copmpositio Math.} \textbf{7}, 283.

\bibitem{SatyaCS} Majumdar SN 2005 \emph{Curr. Sci.} \textbf{88}, 2076.

\bibitem{active} Godec A and Metzler R 2015 \emph{Phys. Rev. E} \textbf{92},
010701(R).

\end{thebibliography}
\end{document}